\documentclass{vgtc}                          

\graphicspath{{figures/}{pictures/}{images/}{./}} 

\usepackage{times}                     

\usepackage{tabu}                      
\usepackage{booktabs}                  
\usepackage{lipsum}                    
\usepackage{mwe}   

\usepackage{mathptmx}                  

\onlineid{0}

\vgtccategory{Research}

\vgtcinsertpkg

\usepackage[table]{xcolor}
\usepackage{soul}
\usepackage{hyperref}

\newcommand{\redacted}[1]{REDACTED}

\newcommand{\secref}[1]{\hyperref[#1]{Sec.~\ref*{#1}}}
\newcommand{\appendixref}[1]{\hyperref[#1]{Appendix~\ref*{#1}}}
\newcommand{\figref}[1]{\hyperref[#1]{Fig.~\ref*{#1}}}
\newcommand{\eqnref}[1]{\hyperref[#1]{Eqn.~\ref*{#1}}}
\newcommand{\tabref}[1]{\hyperref[#1]{Table ~\ref*{#1}}}

\definecolor{quoteColor}{HTML}{002145}

\definecolor{exampleColor}{HTML}{9370DB}

\newcommand{\pxx}[1]
{\textbf{P}$_{\textrm{#1}}$}

\newcommand{\parahead}[1]
{%
  \paraheadd{#1}.
}

\newcommand{\paraheadd}[1]
{%
  \vspace{0.07in}%
  \noindent%
  \textbf{\textit{#1}}%
}

\def\subsubsec#1
{\subsubsection{#1}}

\newcommand{\lillink}[1]{\href{#1}{\textsuperscript{\tiny $\nearrow$}}}

\title{Trade-offs in Data Color Palette Design Tools}

\author{Shiyi He\thanks{e-mail: shiyi.he@utah.edu}\\ %
        \scriptsize University of Utah %
\and Andrew McNutt\thanks{e-mail: andrew.mcnutt@utah.edu}\\ %
     \scriptsize University of Utah }%

\abstract{
Designing a color palette for data requires designers to balance multiple constraints, including accessibility and aesthetics. 
Color palette tools support this process through features including direct manipulation, automated palette generation and evaluation, previews, and so on.
Despite their prominence,  relatively little is known about how these different mechanisms shape design across contexts. 
We conducted an exploratory think-aloud crowd work study with 40 self-identified designers. 
Each participant used one of four palette tools selected to span different interaction modalities to complete a series of accessibility- and aesthetics-oriented design tasks. 
We observed two preliminary patterns. First, tool differences were more pronounced in accessibility-constrained tasks. Second, even when accessibility was not explicitly required, some tools produced more accessibility-friendly palettes and prompted more accessibility-oriented thinking. 
In this tool genre, then, system design shapes outcomes both via built-in functionality, as well as by directing designers' attention toward particular constraints and design considerations.
} 

\keywords{color palettes, accessibility, interaction design}

\begin{document}

\vspace{-1em}
\firstsection{Introduction}

\maketitle



Color palette design for visualization is inherently challenging. Designers need to balance suitability for representing data with a host of design basics including perceptual distinguishability, accessibility, thematic fit, and aesthetic preference. 
Frustratingly, these constraints often conflict: a visually harmonious palette may fail contrast requirements, while a technically accessible one may not fit the intended visual style.

Modern data color palette tools (such as Viz Palette, and Color Buddy~\cite{mcnutt2024mixing})
build upon more familiar color palette design tools~\cite{jalal2015color} by supporting this process via different interaction mechanisms and forms of assistance. Some emphasize direct manipulation, allowing designers to fine-tune colors through incremental exploration of visual harmony and perceptual distinction. 
Others help designers inspect palette performance through visualization previews and color vision deficiency simulations. Some also provide evaluative feedback through accessibility metrics or checks---such as warnings for insufficient contrast.
More automated tools generate candidate palettes or provide repair suggestions, partially shifting design work toward system-proposed alternatives.

These mechanisms reflect different assumptions about how palette design should be supported: whether designers should primarily explore manually based on their own judgment, refine their choices through system-supported feedback, or select from system-generated candidates. 
Yet it remains unclear how different tool support and interaction mechanisms relate to palette outcomes and how they influence designers’ processes across task contexts, particularly when tasks involve varying degrees of design constraint.
\vspace{-1em}

To probe this question, we conduct an exploratory study of how color palette tools' mechanisms shape designers' strategies and outcomes through three color palette construction scenarios that impose different types and degrees of design constraint: accessibility-oriented, thematic, and open-ended design.

\begin{figure*}[t]
  \centering
  \includegraphics[width=\linewidth]{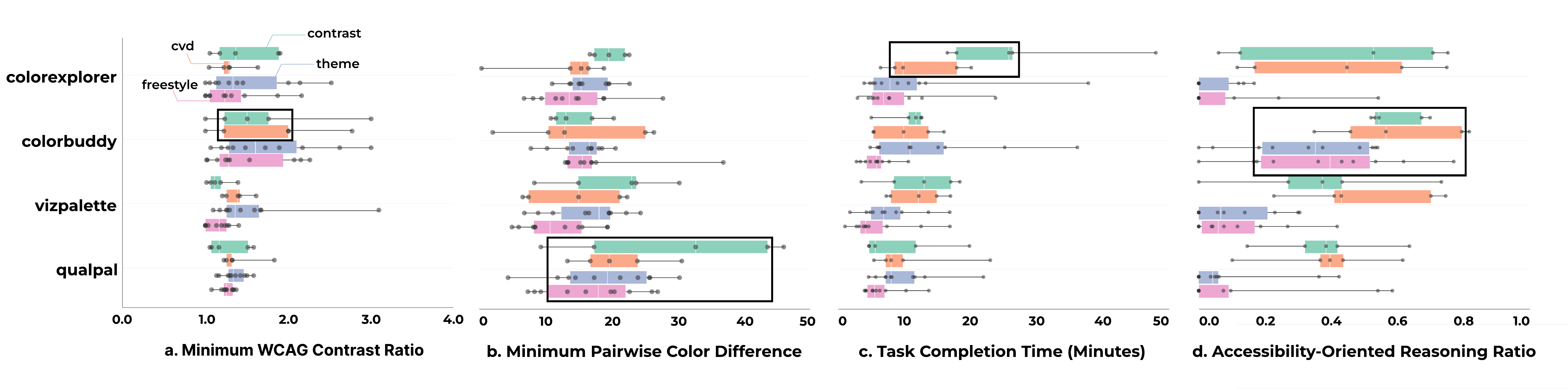}
  \vspace{-2em}
  \caption{Palette outcomes across tools and tasks using minimum WCAG contrast and minimum pairwise $\Delta E_{2000}$ distance. Task completion time with the proportion of thinking time devoted to accessibility-related considerations across tools and tasks. }
  \label{fig:measure}
  \vspace{-2em}
\end{figure*}

\newcommand{\dominant}{Dom.}
\newcommand{\supported}{Supp.}
\begin{table}[t]
\centering
\footnotesize
\label{tab:tools}
\setlength{\tabcolsep}{4pt}
\renewcommand{\arraystretch}{1.12}

\begin{tabular}{
  p{1.7cm}
  p{0.9cm}
  p{0.9cm}
  p{2cm}
  p{1.3cm}
}
\toprule
Tool & Direct\newline Manip. & Auto Gen. & Inspection & Evaluation \\
\midrule
\rowcolor{gray!12}
QualPal\lillink{https://qualpal.cc} & \textit{\textbf{NA}} & \dominant & Viz, CVD  Sim & \supported{} Lint \\
Viz Palette\lillink{https://www.susielu.com/data-viz/viz-palette} & \supported & \textit{\textbf{NA}} & Viz, CVD Sim & \supported{} Lint \\
\rowcolor{gray!12}
ColorExplorer\lillink{https://colorexplorer.com/} & \dominant & \textit{\textbf{NA}} & \textit{\textbf{NA}} & \textit{\textbf{NA}} \\
Color Buddy\lillink{https://color-buddy.netlify.app/} & \dominant & \supported & Viz, CVD Sim & \dominant{} Lint \\
\bottomrule
\end{tabular}
\vspace{1em}

\caption{\scriptsize{
The color palette tools we studied had a range of different interface elements, that were individually dominantly (\dominant), partially (\supported), or not supported. (Viz Palette: \url{https://www.susielu.com/data-viz/viz-palette}; QualPal: \url{https://qualpal.cc}; ColorExplorer: \url{https://colorexplorer.com/}; Color Buddy: \url{https://color-buddy.netlify.app/})
}}
\vspace{-3em}
\end{table}

\vspace{-0.5em}
\section{An Exploratory Think Aloud Study}

We conducted an exploratory between-subject think-aloud study with 40 self-identified designers recruited from Prolific.
Participants were assigned to one of four palette tools which were selected
to span different interaction mechanisms and functionalities, as in \autoref{tab:tools}. 
They completed design tasks with varying constraints (during which they were asked to think aloud): an accessibility-oriented task requiring color vision deficiency-friendly palettes (1a) or sufficient contrast (1b), a thematic constraint task requiring specified brand colors while maintaining visual harmony (2), and an aesthetics-oriented task based on personal preference (3).
See the reVISit-based~\cite{cutler20252025ReVISit} experiment for the task specifics ({\url{https://shiyihe-neko.github.io/reVISit_study/color-buddy/}). 

We evaluated outcome-related metrics, including task completion time across tools and tasks, and the quality of final palettes. 
Palette quality was measured using minimum WCAG contrast ratio between palette colors and the black and white backgrounds to assess foreground-background legibility, and pairwise $\Delta E_{2000}$ distance among palette colors to estimate within-palette separability. 
For the CVD task, we applied the same distance-based criteria within simulated CVD spaces to assess whether palettes remained distinguishable. 
We collected both participants' screen and audio recordings, deductively coding those data using task-derived codes for accessibility concerns, aesthetic preferences, and thematic fit, and to compare how these reasoning patterns varied across tasks and tools. 
Given the exploratory nature of the study, we interpret the results as patterns in outcomes and design processes rather than evidence that one tool outperforms another.


This study had several limitations.
As each tool condition included relatively few participants, limiting our ability to make strong statistical claims, and so its findings are merely gestural or exploratory.
We scoped our tasks around general palette design to observe broad color exploration behaviors, leaving it to future work to explore patterns under data visualization contexts.
While we aimed to select a feature-representative sample of tools, exploring alternative systems---such as ColorBrewer~\cite{harrower2003colorbrewer}, a widely used reference for cartographic and accessible palettes---might reveal different patterns.
Our measures capture only a subset of palette quality, particularly centered on accessibility-related constraints. 
Future work should examine additional quality dimensions, such as perceptual and aesthetic properties. 
Finally, due to a task-assignment misstep, participants completed only one of the two accessibility-oriented task variants (1a versus 1b) rather than both, which led to unequal sample sizes across task variants and made it difficult to disentangle tool effects from task-specific effects.


\vspace{-0.5em}
\section{Findings}
Broadly, we find that color palette tools can shape design outcomes both through their functional mechanisms and through the ways they structure designers' attention and decision-making.

\parahead{Outcome differences between tools were more pronounced in accessibility-oriented tasks}
As in \figref{fig:measure}, palettes created with Color Buddy achieved stronger foreground-background contrast, whereas those created with QualPal showed greater inter-color distinguishability. Participants also completed accessibility-oriented tasks more quickly with both tools than with ColorExplorer.
This pattern suggests that explicit accessibility-support features were most useful when task goals were tied to measurable criteria. 
In QualPal, participants had less direct control over individual color refinement, so they typically selected from automatically generated palettes and used visualization previews and CVD simulations to judge whether the results met the task requirements. Because these features are explicitly designed around measurable accessibility criteria, they naturally guide users toward palettes that perform better on those metrics. 
In the meantime, these functions make relevant constraints easier to identify and address, leading to faster task completion in constrained accessibility tasks. 
In contrast, ColorExplorer relied largely on manual refinement, providing limited cues for evaluating accessibility during palette construction. Together, these examples suggest that tools with explicit accessibility checks made measurable constraints easier to see and act on. This helps participants resolve contrast and CVD-related issues in accessibility-constrained tasks, although it may also steer attention toward what the tool can explicitly evaluate.

\parahead{Tools design shapes outcomes via functionality and attention guidance}
Although the theme and freestyle tasks did not require participants to satisfy accessibility constraints, palettes created with Color Buddy and QualPal still appeared to perform better on accessibility-related measures. 
Specifically, these palettes tended to show higher foreground-background contrast or stronger inter-color separability than those created with other tools (see Figure~\ref{fig:measure}).
The two tools appeared to influence outcomes in different ways. 
In QualPal, accessibility-related outcomes were partly embedded in the generation mechanism itself, as participants initialized the process with their target colors and received automatically generated palettes optimized for distinguishability. As a result, accessibility-related properties were incorporated into the generated palettes regardless of whether accessibility was an explicit task goal. 
In contrast, Color Buddy appeared to influence outcomes by shaping participants' attention during the design process. 
Screen recordings and think-aloud data showed that participants frequently attended to contrast warnings even during theme and freestyle tasks. Several participants repeatedly adjusted colors to clear warnings despite describing thematic fit or visual appeal as their primary objective. Consistent with these observations, participants using Color Buddy exhibited a higher proportion of accessibility-related reasoning across all task types.
These observations suggest that tool mechanisms can shape palette outcomes not only through their built-in functionality, but also by directing designers' attention toward particular design considerations.
While such mechanisms may help satisfy accessibility-related criteria, they can also disproportionately anchor attention, potentially constraining broader design exploration and increasing interaction friction.
For example, some participants expressed frustration with repeatedly resolving accessibility warnings in Color Buddy, while others questioned whether the tool-suggested colors were better than their own. These incidents suggest a potential tradeoff between making design aids visible and actionable and preserving broader design exploration.



\vspace{-0.5em}
\section{Conclusion}

Our exploratory findings suggest that color palette tools not only assist design decisions, but also shape what designers attend to and prioritize.
Tools with explicit accessibility support were most consequential when task goals aligned with measurable criteria such as contrast and CVD-friendly. These mechanisms made accessibility constraints easier to notice and act on, but they may also direct designers' attention toward what the tool can check, suggest, or optimize, potentially drawing attention away from broader creative considerations and increasing interaction friction (which may or may not be useful depending on usage context).

Future studies should investigate when tool support effectively surfaces critical design considerations versus when it prematurely narrows attention to what the system can afford. Additionally, it would be useful to examine how the how different interaction paradigms shape design reasoning, specifically addressing how to balance effective aids with user agency. Understanding these tradeoffs is essential for developing tools that make constraints actionable while preserving creative exploration and interpretation.



\vspace{-0.5em}
\bibliographystyle{abbrv-doi}

\begin{figure*}[t]
  \centering
  \includegraphics[width=\linewidth]{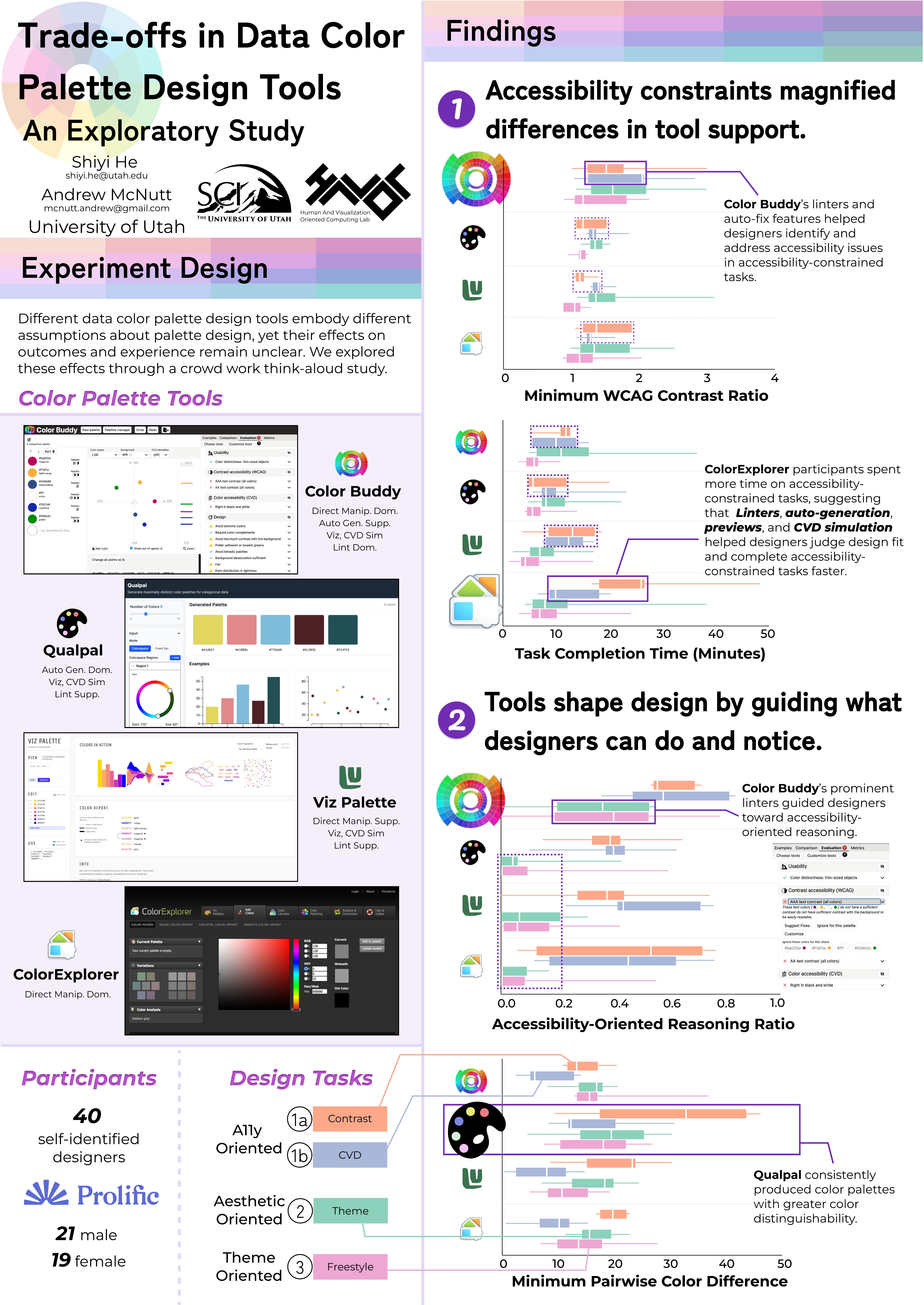}
  \label{fig:poster}
\end{figure*}

\bibliography{template}

@article{cutler20252025ReVISit,
  author  = {Cutler, Zach and Wilburn, Jack and Shrestha, Hilson and Ding, Yiren and Bollen, Brian and Nadib, Khandaker Abrar and He, Tingying and McNutt, Andrew and Harrison, Lane and Lex, Alexander},
  doi     = {10.1109/tvcg.2025.3633896},
  journal = {IEEE TVCG},
  title   = {{Revisit 2: A Full Experiment Life Cycle User Study Framework}},
  year    = {2025}
}

@inproceedings{jalal2015color,
  author    = {Jalal, Ghita and Maudet, Nolwenn and Mackay, Wendy E},
  booktitle = {ACM SIGCHI},
  doi       = {10.1145/2702123.2702173},
  title     = {Color Portraits: From Color Picking to Interacting with Color},
  year      = {2015}
}

@article{mcnutt2024mixing,
  author    = {McNutt, Andrew and Stone, Maureen C and Heer, Jeffrey},
  doi       = {10.1109/tvcg.2024.3456317},
  journal   = {IEEE TVCG},
  title     = {Mixing linters with GUIs: a color palette design probe},
  year      = {2024}
}

@article{harrower2003colorbrewer,
  title={ColorBrewer. org: an online tool for selecting colour schemes for maps},
  author={Harrower, Mark and Brewer, Cynthia A},
  journal={The Cartographic Journal},
  year={2003},
  publisher={Taylor \& Francis}
}
\end{document}